 \documentclass[12pt]{article}  
 \usepackage{graphicx}
 \newcommand{\comma}{\;\; ,}
 \newcommand{\period}{\;\; .}
 \newcommand{\eq}{\; = \;}
 \newcommand{\sep}{\;\; , \;\;}
 \newcommand{\be}{\begin{equation}}
 \newcommand{\bd}{\begin{displaymath}}
 \newcommand{\ee}{\end{equation}}
 \newcommand{\ed}{\end{displaymath}}
 \newcommand{\ba}{\begin{eqnarray}}
 \newcommand{\ea}{\end{eqnarray}}
 \renewcommand{\i}{{\rm i}}
 \newcommand{\e}{{\rm e}}
 \newcommand{\Wb}{\overline{W}}
 \newcounter{storeeqn}
 \newcommand{\up}[1]{\raisebox{0.5ex}{$\scriptstyle #1 $}}
 \newcommand{\Tr}{{\rm Trace} \;}
 \newcommand{\spc}{\; \; \; \; \; \; \; \; \;}
 \newcommand{\cls}{\! \! \! \! \! \!}
 \newcommand{\cn}{{\rm cn}  }
 \newcommand{\sn}{{\rm sn}  }
 \newcommand{\dn}{{\rm dn}  }

 \title{Corner transfer matrices in statistical mechanics}

 \author{R. J. Baxter\\
  {\small Centre for Mathematics and its Applications,
    The Australian National University,}\\{\small Canberra,
    ACT 0200, Australia}\\{\small E-mail: none}}

 \date{{\small November 7th, 2006 }}

\begin{document}

 \maketitle

 \begin{abstract}
 Corner transfer matrices are a useful tool in the statistical 
 mechanics of simple two-dimensinal models. They can be very effective 
 way of obtaining series expansions of unsolved models, and of 
 calculating the order parameters of solved ones. Here we review these
 features and discuss the reason why the method fails to give the 
 order parameter of the chiral Potts model.
  \end{abstract}.

 {\small  PACS numbers 64.60.Cn, 05.50.+q }



 \renewcommand{\theequation}{\arabic{section}.\arabic{equation}}

 \section{Introduction}

 In an early work of the author,\cite{RJB1968} he evaluated 
 numerically a sequence of approximations for the free energy (or 
 entropy) of the monomer-dimer system on the square lattice, using a 
 variational approximation for the eigenvalues and 
 eigenvectors of the row-to-row
 transfer matrix. His interest was to see if the model exhibited a 
 phase transition as the density of dimers increased from zero to the 
 close packed value of one-half. No such transition was observed: the 
 model is critical only at close packing, where it can be solved 
 exactly\cite{Kasteleyn1963}. This was reflected in the poorer 
 convergence of the approximations as close packing is approached.
 
 The result may have been negative, but the work did  lead naturally 
 to the development of the useful concept of ``corner transfer 
 matrices''. 

 These matrices build up the lattice  by rotations about the centre.
 For an infinite lattice they are of infinite dimension, but for
 the zero-field Ising model, one can write them
 in terms of the spinor operators (or Clifford algebra) introduced 
 by Kaufman\cite{Kaufman1949} and diagonalize them exactly.\cite{
 RJB1977,Tsang1979} The results were surprisingly simple, and 
 provided a reasonably direct way of obtaining the spontaneous
 magnetization (the order parameter) of the Ising model, compared 
 with the previous verifications of the Onsager-Kaufman 
 result\cite{OnsKauf1949} by Yang,\cite{Yang1952} Montroll, 
 Potts and Ward\cite{MPW1963} and others.

 It was also realised that the corner transfer matrices satisfy
 certain equations. Again, to give exact results the matrices
 and equations must be infinite-dimensional, but a finite truncation  
 corresponds to a variational approximation of the type
 used for the monomer-dimer system. Such truncations can give 
 useful approximations to the exact results, even for models which 
 have not been solved exactly. For instance,
 Baxter and Enting\cite{BaxEnt1979} were able to obtain 24 terms in 
 the low-temperature 
 series expansion of the Ising model in a field, using only 15 by 15 
 matrices. This was a great improvement on the 12-term series 
 expansions previously obtained.\cite{Sykes1973}

 Here we present these equations for a quite general  
 ``edge-interaction''  model,
 where spins live on sites of the square lattice and the interactions 
 are between adjacent spins. We discuss the simplifications that 
 arise for a ``solvable'' model, i.e. one whose Boltzmann weights 
 satisfy the star-triangle relation. We emphasize the importance
 of the ``rapidity-difference'' property and indicate how it ensures
 that the corner transfer matrices commute and have a very simple
 eigenvalue spectrum. This makes it easy to obtain the order 
 parameters of such models.


 Finally we discuss the chiral Potts model, and indicate how the lack 
 of a rapidity difference property for that model prevents its 
 solution by the same corner transfer matrix techniques.

 For definiteness, here we consider only $Z_N$-invariant models
 (such as the Ising, self-dual Potts and chiral Potts models), where
 each 
 spin can take $N$ values and the Boltzmann weights depend only 
 on the difference (modulo $N$) of the spin values. The extension
 to more general edge-interaction models (perhaps including
 site weight functions as in the solvable Kashiwara-Miwa 
 model\cite{KashMiwa} - \cite{Gaudin}) is straightforward.


 \section{Square lattice edge-interaction models}

\begin{figure}[hbt]

 \begin{picture}(420,210) (-30,0)
 \setlength{\unitlength}{0.28mm}
  \multiput(5,15)(5,0){76}{.}
 \multiput(5,75)(5,0){76}{.}
 \multiput(5,135)(5,0){76}{.}
 \multiput(5,195)(5,0){76}{.}
  \thicklines

 \put (4,12) {\large $< $}
 \put (5,12) {\large $< $}
 \put (6,12) {\large $< $}

 \put (4,72) {\large $< $}
 \put (5,72) {\large $< $}
 \put (6,72) {\large $< $}

 \put (4,132) {\large $< $}
 \put (5,132) {\large $< $}
 \put (6,132) {\large $< $}

 \put (4,192) {\large $< $}
 \put (5,192) {\large $< $}
 \put (6,192) {\large $< $}

 \put (42,230) {\large $\wedge$}
 \put (42,229) {\large $\wedge$}
 \put (42,228) {\large $\wedge$}

 \put (102,230) {\large $\wedge$}
 \put (102,229) {\large $\wedge$}
 \put (102,228) {\large $\wedge$}

 \put (162,230) {\large $\wedge$}
 \put (162,229) {\large $\wedge$}
 \put (162,228) {\large $\wedge$}

 \put (222,230) {\large $\wedge$}
 \put (222,229) {\large $\wedge$}
 \put (222,228) {\large $\wedge$}

 \put (282,230) {\large $\wedge$}
 \put (282,229) {\large $\wedge$}
 \put (282,228) {\large $\wedge$}

 \put (342,230) {\large $\wedge$}
 \put (342,229) {\large $\wedge$}
 \put (342,228) {\large $\wedge$}

 \thinlines


 \put (176,102) {{\large \it a}}
 \put (251,24) {{\large \it i}}
 \put (296,99) {{\large \it j}}
 \put (280,57) {$W(\sigma_i \! - \! \sigma_j)$}
\put (131,24) {{\large  \it k}}
 \put (55,99) {{\large \it l}}
 \put (95,87) {$\overline{W}(\sigma_k \! - \! \sigma_l)$}

 \put (392,13) {{\large \it q}}
 \put (392,73) {{\large \it q}}
 \put (392,133) {{\large $ q' $}}
 \put (392,193) {{\large $ q' $}}

 \put (195,105) {\circle{7}}

 \put (16,45) {\line(1,-1) {60}}
 \put (16,165) {\line(1,-1) {180}}
 \put (76,225) {\line(1,-1) {117}}
 \put (198,103) {\line(1,-1) {117}}
 \put (196,225) {\line(1,-1) {180}}
 \put (316,225) {\line(1,-1) {60}}
 \put (16,165) {\line(1,1) {60}}
 \put (16,45) {\line(1,1) {180}}
 \put (76,-15) {\line(1,1) {117}}
 \put (198,107) {\line(1,1) {118}}
 \put (196,-15) {\line(1,1) {180}}
 \put (316,-15) {\line(1,1) {60}}

 \put (75,105) {\circle*{7}}
 \put (315,105) {\circle*{7}}
 \put (75,-15) {\circle*{7}}
 \put (195,-15) {\circle*{7}}
 \put (315,-15) {\circle*{7}}

 \put (15,45) {\circle*{7}}
 \put (135,45) {\circle*{7}}
 \put (255,45) {\circle*{7}}
 \put (375,45) {\circle*{7}}

 \put (15,165) {\circle*{7}}
 \put (135,165) {\circle*{7}}
 \put (255,165) {\circle*{7}}
 \put (375,165) {\circle*{7}}

 \put (75,225) {\circle*{7}}
 \put (195,225) {\circle*{7}}
 \put (315,225) {\circle*{7}}

 \put (42,-40) {{\large \it p}}
 \put (102,-40) {{\large \it p}}
 \put (162,-40) {{\large \it p}}
 \put (222,-40) {{\large $ p' $}}
 \put (282,-40) {{\large $ p' $}}
 \put (342,-40) {{\large $ p' $}}

 \multiput(45,-25)(0,5){52}{.}
 \multiput(105,-25)(0,5){52}{.}
 \multiput(165,-25)(0,5){52}{.}
 \multiput(225,-25)(0,5){52}{.}
 \multiput(285,-25)(0,5){52}{.}
 \multiput(345,-25)(0,5){52}{.}
 \setlength{\unitlength}{1pt}
 \end{picture}

 \vspace{1.5cm}
 \caption{\footnotesize  The square lattice (solid lines, drawn 
 diagonally), showing a central spin $a$ and the 
 (dotted) rapidity lines.}
 \label{sqlattice}
 \end{figure}
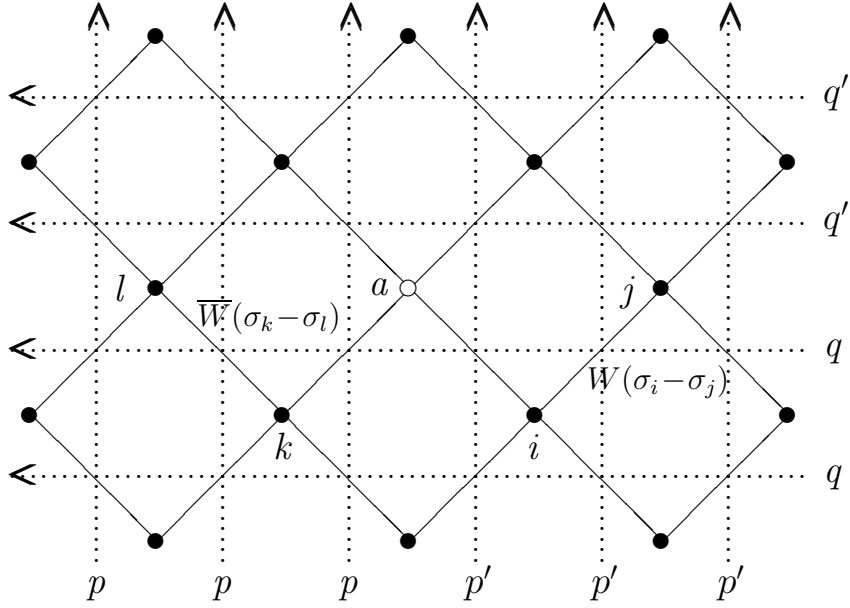

 In Figure \ref{sqlattice} we have drawn the square lattice
 $\cal L$ diagonally, denoting sites by circles and edges by  solid 
 lines. On each site $i$ there is a ``spin'' $\sigma_i$, taking 
 some discrete set of $N$ values, say $0, 1, \ldots , N-1$.
 Spins on adjacent sites $i$ and $j$, with $j$ above and to the right of 
 $i$, interact with Boltzmann weight function 
 $W(\sigma_i - \sigma_j  )$, as indicated. Similarly, spins on sites
 $k$ and $l$, with $l$ above and to the left of $k$, 
 interact with weight function  $\Wb(\sigma_k - \sigma_l  )$.
 The partition function is
 \be Z \eq \sum_{\sigma} \prod_{i,j}  W(\sigma_i - \sigma_j  )
 \prod_{k,l}  \Wb(\sigma_k - \sigma_l  ) \comma \ee
 the sum being over all values of all the spins on the lattice, 
 the first product over all SW - NE edges, the second over all
 SE - NW edges.

 If the lattice has $L$ sites, we expect the limit
 \be \label{kappa}
 \kappa \eq \lim_{L \rightarrow \infty} Z^{1/L} \ee
 to exist, and to be independent of the manner in which $\cal L$
 becomes large, so long as it becomes infinite in all directions.

  In Figure \ref{sqlattice} we have singled out a central site with 
 spin $a$. The average of any function $f(a)$ is defined
 as
 \be \langle f(a) \rangle \eq Z^{-1} \sum_{\sigma} f(a) \prod_{i,j}  
 W(\sigma_i - \sigma_j  )
 \prod_{k,l}  \Wb(\sigma_k - \sigma_l  ) \period \ee
 We also expect any such average to tend to a limit, provided we impose
 suitable  conditions on the values of the boundary spins, and take a limit
 where $a$ becomes infinitely far from any boundary.

 
 \subsection{Corner transfer matrices}

  Here we define the corner transfer matrices (and the associated 
 row and column matrices) for a general anisotropic 
 edge-interaction model.

\begin{figure}[hbt]

 \begin{picture}(420,210) (-30,0)
 \setlength{\unitlength}{0.28mm}
   \multiput(12,105)(8,0){35}{\bf .}
   \multiput(12,105)(8,0){35}{\bf .}
   \multiput(13,135)(8,0){35}{\bf .}
   \multiput(13,135)(8,0){35}{\bf .}

  \multiput(135,-18)(0,8){35}{\bf .}
  \multiput(135,-17)(0,8){35}{\bf .}
  \multiput(165,-18)(0,8){35}{\bf .}
  \multiput(165,-17)(0,8){35}{\bf .}

  \thinlines


 \put (16,45) {\line(1,-1) {60}}

 \put (16,105) {\line(1,-1) {46}}

 \put (92,29) {\line(1,-1) {44}}

 \put (16,165) {\line(1,-1) {34}}

 \put (73,108) {\line(1,-1) {123}}

 \put (16,225) {\line(1,-1) {240}}
 \put (46,255) {\line(1,-1) {240}}

 \put (106,255) {\line(1,-1) {123}}
 \put (250,111) {\line(1,-1) {36}}

 \put (167,255) {\line(1,-1) {47}}
 \put (242,180) {\line(1,-1) {44}}

 \put (226,255) {\line(1,-1) {60}}
 
 \put (16,225) {\line(1,1) {30}}
 \put (16,165) {\line(1,1) {90}}
 \put (16,105) {\line(1,1) {150}}

 \put (16,45) {\line(1,1) {125}}
\put (162,191) {\line(1,1) {64}}

 \put (16,-15) {\line(1,1) {45}}
 \put (91,60) {\line(1,1) {27}}
 \put (136,105) {\line(1,1) {27}}
 \put (186,155) {\line(1,1) {25}}
 \put (241,210) {\line(1,1) {43}}

 \put (76,-15) {\line(1,1) {64}}
 \put (163,72) {\line(1,1) {123}}

 \put (136,-15) {\line(1,1) {150}}
 \put (196,-15) {\line(1,1) {90}}
 \put (256,-15) {\line(1,1) {30}}

 \put(75,45){\makebox(0,0){\large $ A_1$}}
 \put(78,54){\it a}

 \put(150,56){\makebox(0,0){\large $ F_1$}}
 \put(153,65){\it a}

 \put(225,45){\makebox(0,0){\large $ B_2$}}

 \put(237,118){\makebox(0,0){\large $ G_2$}}
 \put(241,124){\it b}

 \put(225,190){\makebox(0,0){\large $ A_3$}}
 \put(228,196){\it b}

 \put(150,178){\makebox(0,0){\large $ F_3$}}
 \put(153,187){\it b}

 \put(75,191){\makebox(0,0){\large $B_4$}}

 \put(61,118){\makebox(0,0){\large $G_4$}}
 \put(65,126){\it a}

 \put(-6,132){$\rho $}

\put(-6,102){$\lambda $}

 \put(131,-37){$\mu $}
 \put(161,-37){$\nu $}
 \put(300,102){$\tau $}

 \put (135,105) {\circle*{7}}
 \put (165,135) {\circle*{7}}
 \put (120,90) {\large \it a}
 \put (173,141) {\large \it b}

 \setlength{\unitlength}{1pt}
 \end{picture}

 \vspace{1.5cm}
 \caption{\footnotesize  The parts of the lattice corresponding to 
the various matrices $A, B, F, G$.}
 \label{ctms}
 \end{figure}
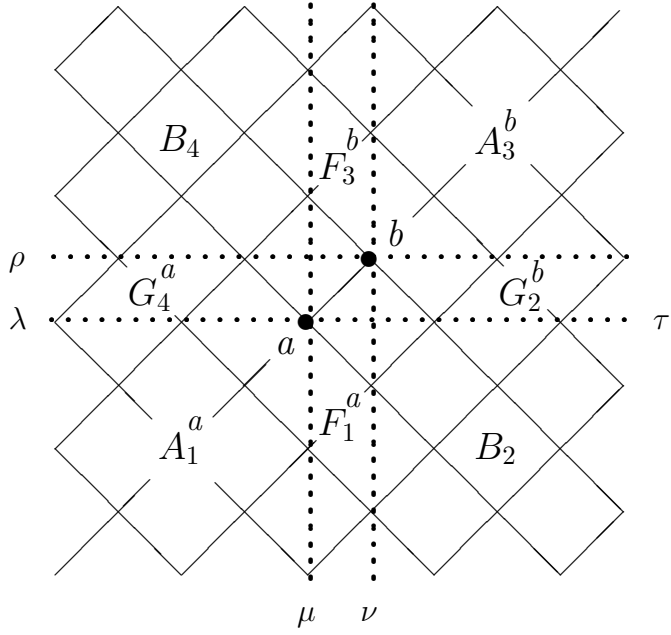

 We have again drawn the diagonal square lattice in Fig. \ref{ctms}
 and have singled out a central edge linking two sites with spins
 $a$ and $b$. We have then used dotted lines to  divide the lattice 
 into nine parts: the 
 central edge, four ``corners'' labelled respectively 
 $A_1^{\up{a}},B_2,A_3^{\up{b}},B_4$, and four columns or 
 rows $F_1^{\up{a}}, G_2^{\up{b}},
 F_3^{\up{b}}, G_4^{\up{a}}$.

 Take all the spins on the outermost boundary of the lattice to have 
 some fixed value,\footnote{For an ordered system, the boundary spins 
 should be set to values consistent with the 
 ground state of the particular phase under consideration.}
 say zero. Let $\lambda$ be the set 
 of spins on the dotted line to the left of
 the spin $a$ and $\mu$ the set of spins on the dotted line below $a$.
 Consider all the edges of the lattice below $\lambda$ and to the left 
 of $\mu$, i.e. all the edges in the lower-left corner of the lattice.
 Take the product over all such edges of the appropriate weight 
 functions
 $W$ or $\Wb$, and sum over all the spins internal to that corner.
 The resulting partition function of the corner will be a function of 
 $a, \lambda$ and $\mu$. Write it as
 $(A_1^{\up{a}} )_{\lambda, \mu}$.


 Similarly, let  $(F_1^{\up{a}} )_{\mu, \nu}$ be the product of the
 weight functions of the edges between $\mu$ and $\nu$. Let 
 $(B_2)_{\nu, \tau}$ be the partition function of the lower-right corner  
 of  the lattice, i.e. the portion to the right of $\mu$ and below 
 $\tau$, and so on. Then the partition function of the complete lattice
 is $Z= Z_W$, where
 \be \label{prod}
 Z_W \eq  \sum W(a-b) (A_1^{\up{a}} )_{\lambda, \mu} 
 (F_1^{\up{a}} )_{\mu, \nu}
 (B_2)_{\nu, \tau} \ldots (G_4^{\up{a}} )_{\rho, \lambda} \comma \ee
 the sum being over all values of $a, b$ and all the spins on the 
 dotted lines of Fig. \ref{ctms}.

 Plainly we can regard $(A_1^{\up{a}} )_{\lambda, \mu}$ as the 
 element $(\lambda, \mu)$ of a matrix $A_1^{\up{a}} $, 
 $(F_1^{\up{a}} )_{\mu, \nu}$ as the element
 $(\mu, \nu)$ of a matrix $F_1^{\up{a}} $, etc. Then (\ref{prod})
 simplifies to 
 \be \label{prod1}
 Z_W \eq  \sum_{a,b} W(a-b) \, \Tr A_1^{\up{a}}  F_1^{\up{a}} 
 B_2 G_2^{\up{b}} A_3^{\up{b}}   F_3^{\up{b}} B_4 G_4^{\up{a}}  
 \comma \ee
 the sum now being only over the spins $a, b$.

 There are other ways of building up $Z$ using the matrices
 $A_i, B_i, F_i, G_i$ as building blocks, most of them being simpler 
 than (\ref{prod1}). They are
 \bd
 Z_{\Wb} \eq  \sum_{a,b} \Wb (a-b) \, \Tr B_1 G_1^{\up{a}} 
  A_2^{\up{a}}  F_2^{\up{a}} 
 B_3 G_3^{\up{b}} A_4^{\up{b}}   F_4^{\up{b}}  \comma \ed

 \bd Z_A \eq \sum_a \, \Tr A_1^{\up{a}} A_2^{\up{a}} A_3^{\up{a}} 
 A_4^{\up{a}} \sep Z_B \eq \Tr B_1 B_2 B_3 B_4 \comma \ed

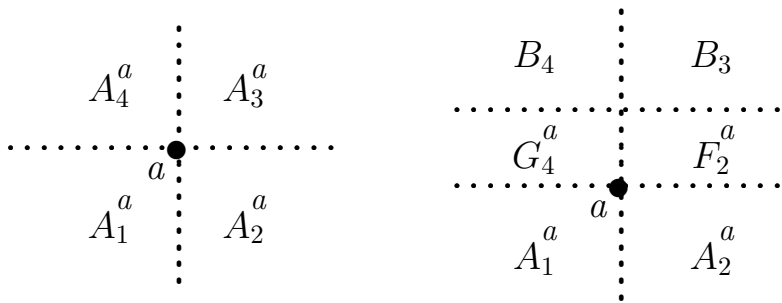
\begin{figure}[hbt]

 \begin{picture}(420,180) (-30,10)
 \setlength{\unitlength}{0.28mm}
   \multiput(12,62)(8,0){20}{\bf .}
   \multiput(12,62)(8,0){20}{\bf .}
  
  \multiput(92,-2)(0,8){16}{\bf .}
  \multiput(92,-1)(0,8){16}{\bf .}
   
  \thinlines

 \put(61,25){\makebox(0,0){\large $ A_1$}}
 \put(64,34){\it a}

 \put(125,25){\makebox(0,0){\large $ A_2$}}
 \put(128,34){\it a}

 \put(61,90){\makebox(0,0){\large $ A_4$}}
 \put(64,96){\it a}

 \put(125,90){\makebox(0,0){\large $ A_3$}}

 \put (93,62) {\circle*{9}}
 \put (78,48) {\large \it a}

 \put(128,96){\it a}


   \multiput(224,44)(8,0){20}{\bf .}
   \multiput(224,44)(8,0){20}{\bf .}

  \multiput(224,80)(8,0){20}{\bf .}
   \multiput(224,80)(8,0){20}{\bf .}

  \multiput(302,-10)(0,8){18}{\bf .}
  \multiput(302,-9)(0,8){18}{\bf .}
   
 \put (303,44) {\circle*{9}}
 \put (288,30) {\large \it a}

 \put(263,10){\makebox(0,0){\large $ A_1$}}
 \put(266,19){\it a}

 \put(347,10){\makebox(0,0){\large $ A_2$}}
 \put(350,19){\it a}

 \put(263,106){\makebox(0,0){\large $ B_4$}}
 \put(347,106){\makebox(0,0){\large $ B_3$}}

 \put(263,58){\makebox(0,0){\large $ G_4$}}
 \put(266,67){\it a}

 \put(347,58){\makebox(0,0){\large $ F_2$}}
 \put(350,67){\it a}

 \setlength{\unitlength}{1pt}
 \end{picture}

 \vspace{1.5cm}
 \caption{ The partition functions $Z_A$, $Z_2$. We have omitted the 
 edges of $\cal L$, but their positions can be deduced from the full 
 circle denoting the lattice spin $a$.}
 \label{Z_A}
 \end{figure}

 \bd Z_i \eq \sum_a \, \Tr A_i^{\up{a}} F_i^{\up{a}} B_{i+1} B_{i+2}
 G_{i+2}^{\up{a}}  A_{i+3}^{\up{a}} \comma \ed
 where $i = 1, \ldots ,4 $ and the suffixes containing $i$ are to be 
 interpreted modulo 4. We have indicated how $Z_A$ and $Z_2$ are 
 constructed in Fig. \ref{Z_A}.

 Note that $F_i^{\up{a}}$ always follows $A_i^{\up{a}}$ and 
 precedes $B_{i+1}$, while  $G_i^{\up{a}}$ follows $B_i$
 and precedes $A_{i+1}^{\up{a}}$.

 If we keep the sizes of the blocks corresponding to the 
 $A_i,B_i,F_i,G_i$ matrices fixed (and mutually consistent), then
 $Z_A, Z_B$ correspond to  the smallest lattices, $Z_W$, $Z_{\Wb}$
 to the largest, and $Z_1, \ldots Z_4$ to intermediate ones. Using
 (\ref{kappa}), we find that in the limit of
 the lattices becoming large, the partition function per site is
 \be \label{varn} 
 \kappa  \eq \frac{Z_A Z_B Z_W Z_{\Wb}}{Z_1 Z_2 Z_3 Z_4 } \period 
 \ee

  We shall find it convenient to define four Boltzmann weight 
 functions $U_1, \ldots, U_4$ by
 \ba U_1(n) = W(n) & , &   U_2(n) = \Wb(n) \comma \nonumber \\
 U_3(n) = W(-n) & , &   U_4(n) = \Wb(-n) \ea
 and to set 
 \be {\tilde{Z}}_1 =  {\tilde{Z}}_3 = Z_W \sep 
 {\tilde{Z}}_2 =  {\tilde{Z}}_4 = Z_{\Wb} \period \ee

 The equations \cite{RJB1981, RJB1982} for the corner transfer matrices
  can be obtained
 by requiring that the expression (\ref{varn}) be stationary with repect
 to variations in the matrices $A_i,B_i,F_i,G_i$. This 
 variational principle leads to the four equations
 \addtocounter{equation}{1}
 \setcounter{storeeqn}{\value{equation}}
 \setcounter{equation}{0}
 \renewcommand{\theequation}
  {\arabic{section}.\arabic{storeeqn}\alph{equation}}

 \be \label{eqq1} \sum_b U_i(a-b) G_{i+1}^{\up{b}} 
 A_{i+2}^{\up{b}} F_{i+2}^{\up{b}}
   B_{i+3} G_{i+3}^{\up{a}} = \eta_i \, B_{i+2} G_{i+2}^{\up{a}} 
 A_{i+3}^{\up{a}} \comma \ee
 \be \label{eqq2}\sum_b U_{i-1}(a-b) 
 F_{i-1}^{\up{a}} B_{i} G_{i}^{\up{b}}
   A_{i+1}^{\up{b}} F_{i+1}^{\up{b}} = \eta'_i \,  
  A_{i}^{\up{a}} F_{i}^{\up{a}}  B_{i+1} \comma \ee
 \be \label{eqq3} F_{i}^{\up{a}}   B_{i+1} B_{i+2}  
    G_{i+2}^{\up{a}} = \xi_i \,  
    A_{i+1}^{\up{a}}   A_{i+2}^{\up{a}} \comma \ee
 \be \label{eqq4}\sum_b G_{i-2}^{\up{b}}    A_{i-1}^{\up{b}}
   A_{i}^{\up{b}} F_{i}^{\up{b}} = \xi'_i \,  
  B_{i-1}   B_{i} \period \ee

 \setcounter{equation}{\value{storeeqn}}
 \renewcommand{\theequation}{\arabic{section}.\arabic{equation}}

 Our notation is now such that rotating the lattice through 
 90{$^\circ$} is equivalent to incrementing $i$ by 1 in all 
 suffixes.

 Pre-multiplying both sides of (\ref{eqq1}) by $ A_{i}^{\up{a}}
 F_{i}^{\up{a}} B_{i+1}$ and summing over $a$, we find that
 \be \eta_i \eq \tilde{Z}_i/Z_i \period \ee
 Similarly,
 \be \eta'_i \eq \tilde{Z}_{i+1}/Z_i  \sep
   \xi_i = Z_{i}/Z_A \sep    \xi'_i = Z_{i}/Z_B   \ee
 and, for instance,
 \be \kappa \eq {\eta_i \, \eta'_{i+2}}/({\xi_{i+1} 
   \, \xi'_{i+3}}) \period \ee

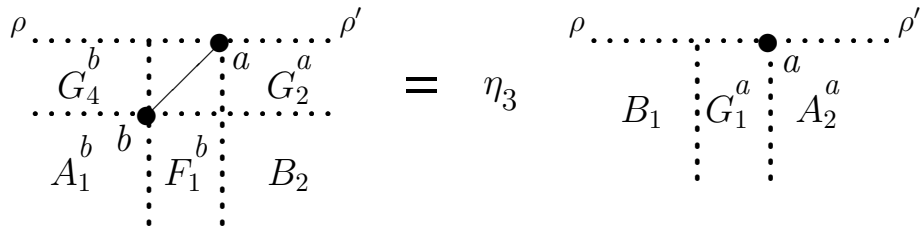
\begin{figure}[hbt]

 \begin{picture}(420,180) (-25,10)
 \setlength{\unitlength}{0.27mm}

   \multiput(10,44)(8,0){19}{\bf .}
   \multiput(10,44)(8,0){19}{\bf .}

  \multiput(10,80)(8,0){19}{\bf .}
   \multiput(10,80)(8,0){19}{\bf .}

  \multiput(102,-10)(0,8){12}{\bf .}
  \multiput(102,-9)(0,8){12}{\bf .}

 \multiput(66,-10)(0,8){12}{\bf .}
  \multiput(66,-9)(0,8){12}{\bf .}

 \put (67,44) {\circle*{9}}
 \put (51,27) {\large \it b}

 \put (103,80) {\circle*{9}}
 \put (108,67) {\large \it a}

 \put (67,44) {\line(1,1) {38}}

 \put(34,58){\makebox(0,0){\large $ G_4$}}
 \put(37,67){\it b}

 \put(30,15){\makebox(0,0){\large $ A_1$}}
 \put(33,24){\it b}

 \put(85,15){\makebox(0,0){\large $ F_1$}}
 \put(90,24){\it b}

 \put(137,15){\makebox(0,0){\large $ B_2$}}

 \put(137,58){\makebox(0,0){\large $ G_2$}}
 \put(140,67){\it a}

 \put (195,57) {\line(1,0) {15}}
 \put (195,63) {\line(1,0) {15}}
 \put (195,56) {\line(1,0) {15}}
 \put (195,62) {\line(1,0) {15}}
 
 \put(233,55){\large $\eta $}

 \put(242,47){$ \textstyle  3 $}

 \put (0,86) {$\rho $}
 \put (162,86) {$\rho ' $}


  \multiput(285,80)(8,0){19}{\bf .}
   \multiput(285,80)(8,0){19}{\bf .}

  \multiput(336,12)(0,8){9}{\bf .}
  \multiput(336,13)(0,8){9}{\bf .}

 \multiput(372,12)(0,8){9}{\bf .}
  \multiput(372,13)(0,8){9}{\bf .}

 \put (373,80) {\circle*{9}}
 \put (379,66) {\large \it a}

 \put (275,86) {$\rho $}
 \put (437,86) {$\rho ' $}

 \put(311,46){\makebox(0,0){\large $ B_1$}}

 \put(353,46){\makebox(0,0){\large $ G_1$}}
 \put(356,55){\it a}

 \put(397,46){\makebox(0,0){\large $ A_2$}}
 \put(400,55){\it a}

 \setlength{\unitlength}{1pt}
 \end{picture}

 \vspace{1.5cm}
 \caption{\footnotesize  Graphical representation of the elements
 $(\rho, \rho')$ of the two sides of equation 
 (\ref{eqq1}) for $i = 3$.}
 \label{1steqn}
 \end{figure}

 The equations (\ref{eqq1}) - (\ref{eqq4}) can be represented
 graphically, each side being a semi-lattice, e.g. for $i = 3$ 
 the equation (\ref{eqq1}) can be represented as in 
 Fig. \ref{1steqn}.

  We can only expect these equations to be exact (for an infinite 
 system) when the matrices are infinitely dimensional. However,
 if we take them to finite-dimensional, then the fact that 
 (\ref{eqq1}) - (\ref{eqq4}) are derived from the variational 
 principle for $\kappa$ ensures that these equations are 
 mutually consistent and that they will define the
  $A_i, B_i, F_i, G_i$ (to within irrelevant similarity 
 transformations). There will be many solutions - a test
 of the utility of any particular solution is that it should
 reproduce as many of the largest eigenvalues of of
 $A_1^{\up{a}} A_2^{\up{a}} A_3^{\up{a}} A_4^{\up{a}}$ and 
 $B_1 B_2 B_3 B_4$ as possible for that truncation.

 Such truncations can provide a powerful tool for performing
 numerical or series expansion calculations on otherwise unsolved
 models.\cite{RJB1999} Indeed, it was such a calculation that led 
 to the exact solution of the hard hexagon 
 model.\cite{BaxTsang1980,RJB1980}


 \section{Solvable models}

  We now focus our attention on models that are known to be solvable
 and satisfy the star-triangle relation. Their corner transfer matrix 
 have special simplifying properties. These properties 
 are true only in the 
 infinite-dimensional limit, but for any given truncation they
 will be approximately true, in particular they will hold to
 appropriate orders in a low-temperature series expansion.

 In general, if a model satisfies the star-triangle relation, then
 its weights $W(n), \Wb(n)$ depend on two variables $p$ and $q$, 
 called  ``rapidities''. The variable $p$ is assocaited with the 
 vertical direction of the lattice $\cal L$, $q$ with the 
 horizontal direction. We display
 this dependence by writing the weights as $W_{pq}(n), \Wb_{pq}(n)$.
 The star-triangle relation relates three sets of Boltzmann weights, 
 with rapidities $(p,q)$, $(p,r)$, $(q,r)$. It is
 \ba \label{startri}
  \sum_d \Wb_{qr}(b-d) \! \! \!   & &  \cls W_{pr}(a-d)\, 
 \Wb_{pq}(d-c)  \spc \spc \spc \nonumber \\
  \spc &&  \eq R_{pqr} \, 
  W_{pq}(a-b) \Wb(b-c) W_{qr}(a-c) \comma \ea
 for all $p, q,  r$ and values of the three external spins 
 $a, b, c$.  The sum is over all values of the internal spin $d$. 

 A second relation must also hold, where all the spin-difference 
 arguments in (\ref{startri}) are negated, so that $b-d$ becomes 
 $d-b$, $a-b$ becomes $b-a$, etc. Provided we impose cyclic
 boundary conditions,  (\ref{startri})
 ensures  that the row-to-row transfer matrices of two models, one 
 with rapidities $(p,q)$, the other with $(p,r)$, 
 commute.\cite{RJB1982} Thus it is natural to allow $q, r$ to vary 
 from row to row of  the lattice $\cal L$, and in Fig. 
 \ref{sqlattice} we have used this freedom to associate two 
 rapidities $q, q'$ with the horizontal rows of the lattice.
 Moreover, this is true even if the 
 rapidities on the dotted vertical lines of Fig. \ref{sqlattice} 
 are different, e.g. if the lines to the left of $a$ have
 rapidity $p$, while those to the right have rapidity $p'$,as  shown.

 More generally, we can allow a different rapidities $p$ for each 
 of the dotted vertical lines in  Fig. \ref{sqlattice} , and a different
 rapidity $q$ for each horizontal line. The Boltzmann
 weight of any edge of $\cal L$ is  $W_{pq}(n)$ or $\Wb_{pq}(n)$,
 where the $p$ and $q$ are the rapidities of the dotted lines
 intersecting that edge.

  It follows  that the eigenvectors of these transfer 
 matrices  are {\em independent } of 
  horizontal rapidities such as the $q, q'$  shown in Fig. 
 \ref{sqlattice}. The correlations of spins lying 
 within a single row of $\cal L$ will therefore also be 
 independent of $q, q'$.

  But (to within an overall irrelevant normalization factor ) 
 the expressions represented by Fig. \ref{1steqn} are  
 precisely such correlations, involving only the spin sets 
 $\rho, \rho'$ and the spin $a$ on the upper edge of each side.  
 This implies that
 \be \label{indep}
 (B_1 G_1^{\up{a}} A_2^{\up{a}})_{\rho,\rho'}  \; = \; \; 
 {\rm independent \; of \; } q, q'  \period \ee

 There is a problem with this argument: we are using fixed-spin 
 boundary 
 conditions, not cyclic ones. However, provided the conditions
 are commensurate with the ground state of the system, we expect
 (\ref{indep}) to be true in the limit when the lattice is large and 
 only a finite number of the spins in  $\rho, \rho'$ differ from the
 ground-state values. This is how we construct the 
 infinite-dimensional matrices, so  we believe (\ref{indep}) to be 
 true in this limit.

  For $i$ odd, it follows that the matrices
  \be  \label{exps}
  A_i^{\up{a}} A_{i+1}^{\up{a}} \sep B_i B_{i+1} \sep
  A_i^{\up{a}} F_i^{\up{a}}  B_{i+1} \sep 
  B_i G_i^{\up{a}} A_{i+1}^{\up{a}} \ee
  are independent of horizontal rapidities such as $q, q'$. 
  Similarly,for $i$ even, they are independent of vertical 
  rapidities such as $p, p'$.
 
  Each matrix will depend only on the rapidities $p, q$ of that
 particular, corner, row or column. For instance $B_1$
 will depend only on the rapidities of the lower-left corner. For 
 each matrix we take all the $p$'s to be that same,and all the $q$'s, 
 and write the matrix as a function of $p$, $q$, e.g. 
 \bd  B_1 \eq B_1(p,q) \period \ed

 For $i$ odd, each of the expressions (\ref{exps}) is a product
 of matrices corresponding to blocks of $\cal L$ that occupy the
 same rows of $\cal L$. The horizontal rapidity $q$ must therefore 
 be the same
 for each matrix within the expression, but the vertical $p$ 
 rapidities may  be different. Thus for instance  the above 
 argument gives
 \bd  B_1(p,q) B_2(p',q) \eq  {\rm independent \; of \; } q  
 \ed
 for all $p, p', q$. Fixing $p'$ and assuming the $B_i$ matrices
 to be invertible, it follows that there exist invertible matrices 
 $R(p), S(q)$ such that
 \bd   B_1(p,q) = R(p) S(q) \period \ed

 This is a factorization property. All the corner matrices
 $A_i^{\up{a}}$, $B_i$ similarly factor and we find that
 \bd A_i^{\up{a}}(p,q) \eq [X_{i-1}^{\up{a}} (\tilde{p}_i)]^{-1}
    {\cal A}_i^{\up{a}} \, X_{i}^{ \up{a}} \! (\tilde{q}_i) 
 \comma \ed
 \be \label{redeqns}
   \cls B_i(p,q) \eq [Y_{i-1} (\tilde{p}_i)]^{-1}
    {\cal B}_i \, Y_{i} (\tilde{q}_i) \comma \ee
 \bd F_i^{\up{a}}(p,q) \eq [X_{i}^{\up{a}} (\tilde{q}_i)]^{-1}
    {\cal F}_i^{\up{a}}(\tilde{p}_i )\,  Y_{i} (\tilde{q}_i) 
 \comma  \ed
 \bd G_i^{\up{a}}(p,q) \eq [Y_{i} (\tilde{q}_i)]^{-1}
    {\cal G}_i^{\up{a}}(\tilde{p}_i ) \, X_{i}^{\up{a}} \! 
 (\tilde{q}_i)  \comma \ed
 where $\tilde{p}_i = p, \tilde{q}_i = q$ if $i$ is odd; and
 $\tilde{p}_{i+1} = \tilde{q}_i$,  $\tilde{q}_{i+1} = 
 \tilde{p}_i$ for all $i$. Note that the matrices 
 ${\cal A}_i^{\up{a}}$, ${\cal B}_i$ are independent of 
 $p$ and $q$, while ${\cal F}_i^{\up{a}}$, ${\cal G}_i^{\up{a}}$
 depend only on the single rapidity ${\tilde{p}}_i$.

 The choice of  the $X_i, Y_i$ is not unique: each can be 
 pre-multiplied by different constant invertible matrices
 $C_i, D_i$ and the ${\cal A}_i$, ${\cal B}_i$, ${\cal F}_i$,
 ${\cal G}_i$ then adjusted accordingly so as to ensure that
 (\ref{redeqns}) remains satisfied. However, this does
 {\em not} change the eigenvalues of 
 \bd
   {\cal M}^{\up{a}} \eq {\cal A}_1^{\up{a}} {\cal A}_2^{\up{a}} 
    {\cal A}_3^{\up{a}} {\cal A}_4^{\up{a}} \sep
   {\cal N} \eq {\cal A}_1 {\cal A}_2 
    {\cal A}_3 {\cal A}_4 \period \ed
  These matrices define $Z_A$ and $Z_B$ and play an imortant 
 role in the calculations: 
 for instance, the order parameters are
 \be \label{order}
  \langle \omega^{j a} \rangle \eq \sum_a \omega^{j a} \, \Tr   
 {\cal M}^{\up{a}} \left/\sum_a \Tr {\cal M}^{\up{a}} 
 \right.  \ee
 where $\omega  = \e^{2 \i \pi/N}$ and $j = 1, \ldots , N-1$.

 Substituting these forms into the equations (\ref{eqq1}) - 
 (\ref{eqq4}), all the $X_i$, $Y_i$ matrices cancel out, so the
 effect is to replace the matrices $A_i, B_i, F_i, G_i$ by
 ${\cal A}_i, {\cal B}_i, {\cal F}_i, {\cal G}_i$, where 
 ${\cal A}_i, {\cal B}_i$ are rapidity-independent and
 ${\cal F}_i, {\cal G}_i$ depend only on the single rapidity 
 $\tilde{p}_i$.

 \subsection{Properties and symmetries}

  There are various properties and symmetries that relate
 the corner transfer matrices. For the Ising, self-dual Potts 
 and chiral Potts models\cite{BPAuY1988} it is  true that
 \be W_{pp}(n) \eq 1 \sep \Wb_{pp}(n) \eq \delta_{n,0} \ee
 and there exist rotation and reflection transformations
 $p \rightarrow Rp $, $p \rightarrow Sp $, such that
 \be \label{rotn}
  \Wb_{pq}(n) \eq   W_{q,Rp}(n) \sep   W_{pq}(n) \eq   
  \Wb_{q,Rp}(-n) \comma \ee
  \be
  W_{pq}(n) \eq   W_{Sq,Sp}(n) \sep   \Wb_{pq}(n) \eq   
  \Wb_{Sq,Sp}(-n) \period \ee

 If we write $U_i(n)$ more explicitly as $U_i(p,q|n)$, it follows 
 that $U_i(q,Rp|n) = U_{i+1}(p,q|n)$ and 
 \be A_i^{\up{a}} (p,p) \eq B_i(p,p) \eq {\rm {\bf 1}} 
 {\rm \; \; \; for \; }i = 1 {\rm \, and \, } 3 \comma  \ee
 \be  \label{A1A2}
 A_{i}^{\up{a}} (q,Rp) \eq A_{i+1}^{\up{a}} (p,q) \; , \ldots ,
 \;  G_{i}^{\up{a}} (q,Rp) \eq  G_{i+1}^{\up{a}} (p,q) \comma \ee
 and, for instance, 
 \be  B_i(Sq,Sp) \eq B_{6-i}(p,q)^T \period \ee
 Replacing $p,q$ by $q,Rp$ is simply equivalent to rotating
 the lattice through $90^{\circ}$, i.e. incrementing the suffixes
 $i$ by one.

 These equations impose many restrictions of the matrices
 $X_i, Y_i$, ${\cal A}_i$, ${\cal B}_i$, ${\cal F}_i$, ${\cal G}_i$.
 We shall not explore all of them here, but note that they do imply
 that there exists a matrix function $X^{\up{a}}(p)$ such that we 
 can choose
 \be
  X_1^{\up{a}}(p) = X_4^{\up{a}}(p) = X^{\up{a}}(p) \sep
  X_2^{\up{a}}(p) = X_3^{\up{a}}(p) = \left( {\cal A}_2^{\up{a}} 
 \right) ^{-1} X^{\up{a}}(p) 
 \comma \ee
 \bd  {\cal A}_1^{\up{a}} \eq  {\cal A}_3^{\up{a}} \eq  {\rm {\bf 1}} 
 \ed
 \be \label{AXX}  A_1^{\up{a}}(p,q) \eq \left[ X^{\up{a}}(p) 
 \right] ^{-1} 
   X^{\up{a}}(q) \sep A_2^{\up{a}}(p,q) \eq \left[ X^{\up{a}}(q) 
 \right] ^{-1} 
   X^{\up{a}}(Rp) \comma \ee
 \bd  A_3^{\up{a}}(p,q) \eq \left[ X^{\up{a}}(Rp) \right] ^{-1} 
   X^{\up{a}}(Rq) \sep A_4^{\up{a}}(p,q) \eq \left[ X^{\up{a}}(Rq) 
 \right] ^{-1} 
   X^{\up{a}}(R^2 p) \comma \ed
 where 
 \be  X^{\up{a}}(R^2 p) \eq {\cal M}^{\up{a}}  X^{\up{a}}(p) 
 \period \ee
 Corresponding equations hold for the matrices 
 $Y_i, {\cal B}_i, B_i, {\cal N}_i$, with no superfixes $a$.


\section{Ising model}

 We have seen how the star-triangle relation implies that row-to-row
 transfer matrices commute and corner transfer matrices factor.
 To proceed further we need an important property, namely that for
 many models we can choose the rapidity variables $p, q$ so that
 \be \label{rapdiff}
 W_{pq}(n),  \Wb_{pq}(n) \eq \; {\rm functions \; only \; of \; } 
 q \! -\! p \; {\rm and }\;  n \period \ee
 We call this the {\em rapidity difference} property. It is 
 intimately connected with the fact that for such models the 
 Boltzmann weights 
 live on algebraic curves of genus 0 or 1, so can be naturally
 be parametrized in terms of Jacobi elliptic functions. One had
 become so used to this property that it was taken for granted. It
 was the discovery of the chiral Potts model\cite{BPAuY1988}, 
 which does {\em not} have this property,
 that made its importance obvious.

   The Ising model is a  two-state model (i.e. $N=2$) of the type 
 discussed above. Taking $W_{pq}(0) = \Wb_{pq}(0) = 1$, 
 its other Boltzmann weights are\cite{RJB1988}
 \be \label{WWb}
 W_{pq}(1) = \e^{-2J} =  k' \, {\rm scd}(K-u) \comma \ee
 \be \Wb_{pq}(1) = \e^{-2\overline{J}} =  k' \, {\rm scd} (u)  \comma 
 \ee
 where 
 \be \label{diff}
 u = u_q - u_p \comma \ee
 and the function $\rm{scd} (u)$ is defined by
 \bd {\rm scd}(u) \eq \sn (u/2)/[\cn (u/2) \dn (u/2) ] \comma \ed
 $\sn \, u , \cn \, u, \dn \, u $ being the Jacobi ellptic functions 
 of argument $u$ and modulus $k$.

 Then, setting $k' = (1-k^2)^{1/2} $, we have
 \be \sinh 2 J \eq \frac{\sn \, u }{\cn \, u } \sep
 \sinh 2 \overline{J} \eq \frac{\cn \, u }{k' \sn \, u }  \ee
 and $\sinh 2 J  \sinh 2 \overline{J} = 1/k'$. The conjugate modulus
 $k'$ is small at low temperatures, increasing to one at criticality.
 Within this range the model has ferromagnetic order. The Boltzmann 
 weights satisfy (\ref{rotn}) if we define
 \be u_{Rp} = u_p +K \comma \ee
 they are real and positive provided $u = u_q - u_p$ is real and
 \be 0 <  u_q - u_p < K \comma \ee
 $K, K'$ being the usual complete elliptic integrals. Note that
 this restriction remains satisfied if we replace $p,q$ by $Rq, p$, 
 when $u_q - u_p$ becomes $K-u_q+u_p$.

 The relation (\ref{diff}) manifests the difference property. It has 
 far-ranging consequences, in fact it enables us to calculate the 
 spontaneous magnetization, i.e. the order parameter (\ref{order})
 with $j=1$.\cite[chapter 13]{RJB1982}

 To see this, note that we write the first equation (\ref{AXX}) as
 \be A_1^{\up{a}}(u_q-u_p) \eq \left[ X(u_p)^{\up{a}} 
 \right] ^{-1} X(u_q)^{\up{a}} \period \ee
 Replacing $p,q$ by $q,r$ and multiplying, this gives
 \bd A_1^{\up{a}}(u_q-u_p) A_1^{\up{a}}(u_r-u_q) \eq A_1^{\up{a}}
 (u_r-u_p) \ed
 i.e.
 \be \label{u+v}
 A_1^{\up{a}}(u) A_1^{\up{a}}(v) \eq A_1^{\up{a}}(u+v) \comma \ee
 for arbitrary $u,v$. Interchanging $u$, $v$, it follows that
 \be  A_1^{\up{a}}(u) A_1^{\up{a}}(v) \eq
  A_1^{\up{a}}(v) A_1^{\up{a}}(u) \comma \ee
 so $ A_1^{\up{a}}(u), A_1^{\up{a}}(v)$ {\em commute}, for all 
 $u,v$.

    There is therefore a similarity transformation, independent
 of $u$, that diagonalises $ A_1^{\up{a}}(u)$. Let us assume this 
 transformation has been applied and take  $A_1^{\up{a}}(u)$
 to be diagonal. Then it follows from (\ref{u+v}) that
 each diagonal element of $ A_1^{\up{a}}(u)$ is an {\em exponential}
 function of $u$, i.e.
 \be \label{eig}
 [A_1^{\up{a}}(u)]_{i,i} \eq \exp ( - n_i u) \comma \ee
 where $n_i$ is some number.

  Allow $u = u_q - u_p$ to be an arbitrary complex number. Since
 \bd {\rm scd} (u+2 \i K') = - \, {\rm scd} (u) \comma \ed
 it follows from (\ref{WWb}) that 
 incrementing $u$ by $2 \i K'$ merely negates $W_{pq}(1)$ 
 and $\Wb_{pq}(1)$. From series expansions, we expect  
 $ A_1^{\up{a}}(u)$, and indeed 
 all the functions we have defined in the thermodynamic limit
 of a large lattice, to be an analytic function of $u$ within
 the extended physical domain $0 < \Re (u) < K$, which implies
 \be 
 A_1^{\up{a}}(u+2 \i K' ) \eq (-1)^a \, A_1^{\up{a}}(u ) 
 \period \ee
 It follows that
 \be \label{ni} n_i \eq \pi m_i/(2 K') \comma \ee
 where $m_i$ is an even integer for $a=0$, an odd integer for 
 $a = 1$.

 If we take the low-temperature limit $k' \rightarrow 0$, keeping
 $u$ in the vicinity of $K/2$, then to leading order in
 $q' = \e^{-\pi K/K'}$,
 \be W_{pq}(1) \eq \e^{-\pi u/(2 K')} \sep
 \Wb_{pq}(1) \eq \e^{\pi (u-K)/(2 K')}  \period \ee
 We find that  $A_1^{\up{a}}(u ) $, as 
 originally defined before (\ref{prod}), is already diagonal to 
 leading order. Let the spins to the left of and including
  $a$ in Fig. \ref{ctms}
  be $\ldots , \lambda_2, \lambda_1, a$, where $\lambda_1$ is 
 next to $a$,  $\lambda_2$ is next to $\lambda_1$, etc. Then
 the corresponding diagonal element of  $A_1^{\up{a}}(u ) $
 is indeed of the form (\ref{eig}), (\ref{ni}), the index $i$ 
 being replaced by the spin set $\lambda, $i.e.
 \be \label{resA1}
 A_1^{\up{a}}(u )_{\lambda, \lambda} \eq \exp \{ -\pi\,  u 
   \, m_{\lambda}/(2 K') \} \comma \ee
 where
 \be \label{defm}
  m_{\lambda} \eq 
 |a-\lambda_1 | + 3 |\lambda_1-\lambda_2 | 
  + 5 |\lambda_2-\lambda_3 | + \cdots \period \ee
 We have obtained this result in the low-temperature limit
 $k' \rightarrow 1$. However, for all $k'$ the $m_{\lambda}$
 must be integers and we do not expect them 
 to change discontinuously as $k'$ increases to unity. Hence
 we expect the formulae (\ref{resA1}), (\ref{defm}) for 
 the diagonal elements of  $A_1^{\up{a}}(u )$
 to be {\em exact} for $0 <k' <1$.

 Set $u_1 = u_2 = u_q-u_p$ and $u_2 = u_4 = K-u_q +u_p$. Then 
 from (\ref{A1A2}), 
 \be
 A_i^{\up{a}}(p,q) \eq A_1^{\up{a}}(u_i )  \comma \ee
 so these matrices, together with the product
 \be M^{\up{a}} \eq  A_1^{\up{a}}(p,q)  A_2^{\up{a}}(p,q) 
 A_3^{\up{a}}(p,q)   A_4^{\up{a}}(p,q) \ee
 are diagonalised by the same transformation that diagonalizes
 $A_1^{\up{a}}(u )$, and 
 \be \label{resM}
 M^{\up{a}}_{\lambda, \lambda} \eq x^{m_{\lambda}}  \comma  \ee
 where $x = q' = \e^{-\pi K/K'}$.

 These are also the elements of the diagonalized matrix 
 ${\cal M}^{\up{a}}$: we see that our calculation is consistent 
 with the independence
 of ${\cal M}^{\up{a}}$ on the rapidity variables $p, q$.

 It is now straightforward to calculate the spontaneous magnetization
 from (\ref{order}). First set $\lambda_0 =  a$ and
 \be
 l_i = |\lambda_i - \lambda_{i+1}| \sep i \geq 0 \comma \ee
 so $l_i = 0$ if $\lambda_i= \lambda_{i+1}$, otherwise $ l_i= 1$.
 Take $\lambda_i = 0$ for sufficiently large $i$, then to modulo
 2, 
 \bd a \eq \lambda_0 + \lambda_1 + \lambda_2 + \cdots \period \ed
 and \bd m_{\lambda} \eq \lambda_0 + 3 \lambda_1 + 5 
 \lambda_2 +  \cdots  \period \ed
 {From}  (\ref{order}) and (\ref{resM}),
 \be \langle (-1)^a \rangle \eq \sum_{l} (-1)^a x^{ m_{\lambda}} \left/ 
 \sum_{l} x^{ m_{\lambda}} \right. \period  \ee

 Here $a, m_{\lambda}$ are the linear expressions immediately above
 and the sums are over $l_0, l_1, l_2, \ldots = 0, 1$, without 
 restriction. Each sum therefore factors into a product of individual
 sums over $l_0, l_1, l_2, \ldots$, giving
  \be \langle (-1)^a \rangle \eq \frac{(1-x) (1-x^3) (1-x^5) \cdots }
 {(1+x) (1+x^3) (1+x^5) \cdots }  \period \ee
 Using known elliptic function formulae,\cite[8.197.4]{GR} and 
 \cite[15.1.4b]{RJB1982}  the 
 RHS is $k^{1/4}$, so 
 \be \langle (-1)^a \rangle \eq k^{1/4} \eq (1-{k'}^2)^{1/8} \ee
 which is the famous result announced by Onsager in 
 1949\cite{OnsKauf1949} and derived by Yang in 1952.\cite{Yang1952}


 \section{Summary}

    We have expicitly shown how the star-triangle relation implies that
 the corner transfer matrices factor, causing the reduced matrix 
 functions ${\cal A}_i^{\up{a}},{\cal B}_i,{\cal F}_i^{\up{a}},
 {\cal G}_i^{\up{a}}$ to depend on at most one of the two rapidity
 variables. The rotation (and reflection) relations enable one to 
 express the matrices, for $i = 1, \ldots, 4$, in terms of the
 single $i=1$ case.

 Curiously, this by itself does not seem enough to enable one to 
 calculate quantities such as the order parameters: if one also has 
 the rapidity difference property (\ref{rapdiff}), then the corner 
 transfer matrices commute and can be simultaneously  reduced to
 diagonal form. The eigenvalues are necessarily exponentials in
  $u = u_q-u_p$ and are periodic of period $2\i K'$: this means
 they are defined by certain integers $m_{\lambda}$, and these
 integers can be obtained from a low-temperature limit. The order 
 parameters follow immediately.

 It still seems strange that the method should fail for a model such as
 the chiral Potts model, which does satisfy the star-triangle relation, 
 so in that sense is ``solvable''. As yet the method has only 
 yielded finite series expansions.\cite{RJB1991,RJB1993} 

 The order parameters have now
 been obtained by a specialization of the Jimbo, Miwa, Nakayashiki 
 method,\cite{Jimbo1993} - \cite{RJB2005b} but it would be 
 interesting to rescue the corner transfer matrix technique for 
 this model. For the models with the difference property, 
 diagonalizing the matrices leads directly to the Jacobi elliptic
 parametrization. For the chiral Potts model, is there any way of 
 expressing the matrices that leads to an analogous
 parametrization?


 \smallskip

 \end{document}